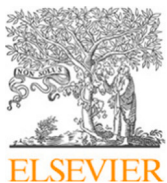
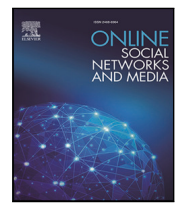

# Information flow estimation: A study of news on Twitter

Tobin South [a,b,c,*], Bridget Smart [a,b], Matthew Roughan [a,b], Lewis Mitchell [a,b]

[a] *School of Mathematical Sciences, University of Adelaide, Adelaide, Australia*
[b] *ARC Centre of Excellence for Mathematical & Statistical Frontiers, Australia*
[c] *MIT Media Lab, Massachusetts Institute of Technology, USA*

A B S T R A C T

News media has long been an ecosystem of creation, reproduction, and critique, where news outlets report on current events and add commentary to ongoing stories. Understanding the dynamics of news information creation and dispersion is important to accurately ascribe credit to influential work and understand how societal narratives develop. These dynamics can be modelled through a combination of information-theoretic natural language processing and networks; and can be parameterised using large quantities of textual data. However, it is challenging to see "the wood for the trees", *i.e.,* to detect small but important flows of information in a sea of noise. Here we develop new comparative techniques to estimate temporal information flow between pairs of text producers. Using both simulated and real text data we compare the reliability and sensitivity of methods for estimating textual information flow, showing that a metric that normalises by local neighbourhood structure provides a robust estimate of information flow in large networks. We apply this metric to a large corpus of news organisations on Twitter and demonstrate its usefulness in identifying influence within an information ecosystem, finding that average information contribution to the network is not correlated with the number of followers or the number of tweets. This suggests that small local organisations and right-wing organisations which have lower average follower counts still contribute significant information to the ecosystem. Further, the methods are applied to smaller full-text datasets of specific news events across news sites and Russian troll accounts on Twitter. The information flow estimation reveals and quantifies features of how these events develop and the role of groups of trolls in setting disinformation narratives. In summary, this work provides a new methodology for examining the information transmitted between content producers in any connected system of natural language, a toolkit with applications to the many networked discourses of our online world.

## 1. Introduction

News plays a fundamental role in society through both its observation of events and its influence on them. The rise of digital news and its corresponding data has allowed for considerably more quantitative analysis [1,2], and yet the dynamics of information transmission between news organisations in the greater news ecosystem is not well understood. By analysing the dynamics of information flow between news organisations, we can better understand the ecosystem of public discourse and the timeliness of news reporting.

The mechanisms of information flow have been seen to impact the spread of true and false news through social media [3], the spread of messages in organisations [4], and the information flow within research and development laboratories [5]. While valuable, many of these models and empirical studies share a common theme: they focus on the transmission and diffusion of singular ideas, or packets of information.

On Twitter specifically, there is a large body of research studying information flows in a variety of contexts, that defies a brief summary. However, it is fair to say that information flows on Twitter are very often modelled by the spread of hashtags or keywords [6–8]. This is attractive from the point of view of modelling, as it makes information flow dynamics amenable to analysis by contagion models, for which a deep literature exists [9]. Unfortunately, keyword- or hashtag-based analyses necessarily discard much of the data in a tweet, namely, the rest of the content outside of the hashtags or keywords in question. The information-theoretic approaches we deploy here aim to utilise more of this data when estimating information flows.

There are some studies using information-theoretic approaches applied to various Twitter data, including time series data on tweet volumes [10], sequences of event times [11], and topic model representations of tweets [12]. Also, not all modelling approaches fall under the contagion modelling framework, with some agent-based approaches being deployed [13]. However, these again model the movement of hashtags rather than full language.

More general information flows can be understood through statistical and information-theoretic measures of information flow [10,14–17]. Several models have been proposed to understand this information flow in networks [18–21]. Here we adapt these approaches to analysing the news ecosystem. The information flow between news organisations is vital to the public discourse and the timeliness of news reporting is highly valued. Social media exemplifies this, where news is often broken and spread virally through the social media ecosystem as stories unfold [22,23], often without reliable verification of information [24].

The news ecosystem is challenging, particularly as observed through social media. Tweets are short, the volume of data is large, and the meaningful data might be small within this large corpus, which also contains conversations, memes, advertising, and spam, from both human and non-human actors. Any measure used to analyse such data needs to be both sensitive and robust. In particular, it needs to be able to compare both large and small news outlets. One approach to aid relative comparisons is normalisation, but (i) is it not obvious that normalised metrics are indeed superior, and (ii) there are many different approaches to normalisation in a network context. Therefore, the first step in this work is to assess seven different normalised and unnormalised metrics for network information flow. The results show that a new metric that normalises by local neighbourhood structure provides a robust estimate of information flow in large networks. The metric's performance, as measured by Pearson correlation coefficient, is 0.97 as compared to the next best at 0.94 and naive estimators at 0.05.

The second major component of the work is an application of these measures to real datasets. First, we present an analysis of a large Twitter dataset to examine the relationships between news outlets over 2019. This analysis confirms a variety of results that are intuitive at least in hindsight, *e.g.,* the largest flow is from the Wall Street Journal to its opinion column WSJOpinion; and Defence One is a significant net information "absorber", because it concentrates on a niche area that is the focus of no other outlets and thus little of its content is replicated. But some conclusions are surprising. For instance, average information contribution to the network is not correlated with the number of followers or the number of tweets. This provides some quantitative evidence supporting the claim that small local organisations and right-wing organisations which have lower average follower counts still contribute significant information to the ecosystem.

Second, we analyse full-text articles from major news organisations are collected for three specific news events, and the measures are applied to analyse how credit can be ascribed and quantified in each case.

Third, the measure is applied to a large corpus of tweets from Russian linked trolls from the Internet Research Agency during the 2016 election to identify which groups of accounts are driving the actively-spread disinformation narratives. This analysis reveals that information generally flows out of accounts posing as Right-leaning authentic accounts, particularly to accounts labelled as `LeftTroll` and `HashtagGamer`. A net information flow is also observed from accounts posing as news sources to accounts posing as commercial users. The median net information flow is significantly different between groups, highlighting the usefulness of our measure to characterise bot account behaviour online. Our results reveal the network of complex interaction dynamics being simulated by this inauthentic group of accounts, and are suggestive of underlying online social influence strategies being deployed by the Internet Research Agency.

The main contributions of this work are:

- The introduction and testing of new measures of information flow that use information-theoretic approaches to identify direction and magnitude of text-based influence in networks of text producers.

- An examination of the sensitivity and reliability of approaches to estimating information flow in the presence of noisy textual data resulting in a new normalised information-flow metric.
- A large, curated dataset of news organisations on Twitter and all their tweets over the 2019 calendar year, as well as the full-text of a subset of articles relating to specific news events.
- An application of the new measures to study the information flow relationships between news organisations – both at a full-year timescale and for specific news events – and an application to examine how information flow relates to behaviour patterns to identify which trolls drove disinformation narratives in the Russian campaign during the 2016 election.

## 2. Background

### 2.1. Measuring information flow

Consider a network in which each node produces segments of natural language text at discrete (possibly random) points in time. Simple measures of entropy based on frequency [25] are insufficient to describe the complexity of the overall text producer, as information is present in both the order of words and possibly the order in which each segment of text is produced. A better measure of this information content is the entropy rate estimator,

$$h(\mathcal{X}) = \frac{N \log N}{\sum_{i=0}^{N} \Lambda_i}, \quad (1)$$

for large $N$, where $\Lambda_i$ is the length of the longest subsequence starting at position $i$ that appear as a contiguous subsequence in the previous $i$ symbols and $N$ is the length of the data. First introduced by Kontoyiannis et al. [26] to estimate the entropy rate of a sequence of text this estimator has been used to find the complexity of sequences of movement patterns [27] and social media predictability [17].

This estimator can then be generalised to create a temporally-aware cross-entropy metric [17,28]. The time-synced cross-entropy rate,

$$h(\mathcal{T} \| S) = \frac{N_\mathcal{T} \log_2 N_S}{\sum_{i=1}^{N_\mathcal{T}} \Lambda_i(\mathcal{T} | S_{\leq t(T_i)})}, \quad (2)$$

is calculated by finding, $\Lambda_i(\mathcal{T} | S_{\leq t(T_i)})$, the longest subsequence starting at position $i$ in the *target* $\mathcal{T}$ that appears as a contiguous subsequence in, $S_{\leq t(T_i)}$, the section of the source *source S* that was created before the text at $i$ was created. In its original context, this meant that the match-length of the text from the current tweet in the target was only calculated against tweets that were created by the source before the time of target tweet. In effect, this ensures that information flow is only measured as it flows forward in time (*e.g.,* if two node tweet the exact same sequence of words, the information will be recorded as having flowed from the node who tweets first to the node who tweeted the second). A diagram visually representing the flow of these $\Lambda_i$'s between sequences and their role in estimation can be seen in Fig. 1.

Note that although this metric provides a notion of direction, it is predicated on a single pair of information sources and lacks some qualities needed in a network setting. Notably, it lacks symmetry and normalisation between targets. In this work we adapt and extend (2) by combining it in a variety of ways to create new metrics that have these desired properties. We achieve this by comparing naive difference measures between these simple entropy rate calculations against more complex compositions of the entropy estimates that combine the estimates with the implicit local network structure or entropy rates calculated on isolated processes.

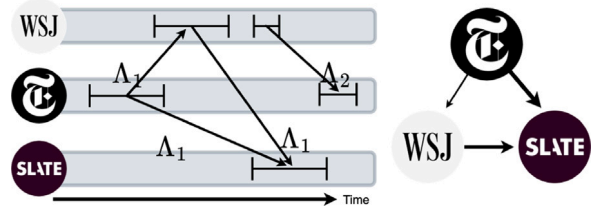

**Fig. 1.** A visual depiction of quoting occurring between three news organisations text streams (left), and the resulting information flow network (right).

## 2.2. Quoter model

The quoter model was introduced to simulate the dynamics of information flow on networks [29] between online friends. It was designed to mimic the information generation process of accounts on social media, where an account (node) creates content by either adding new information to the platform, or copying/quoting information already seen in their feed. We use it here to assess the validity and robustness of information flow metrics in the context of a news-network ecosystem where a similar process is taking place between news outlets.

The quoter model presupposes a network of actors connected by directed edges. These edges indicate that a quoting process is occurring from a source $j$, to a target, $i$. Each such edge is assigned a quote probability, $q_{ij}$, and each node has a self-generation probability, $q_{ii}$, normalised so that $\sum_{\forall i} q_{ij} = 1$. These probabilities are used to decide how a node behaves at each time step of a simulation.

The model repeatedly performs a process of text generation for $T$ steps. At each time point $t$, every node creates text through one of two processes.

1. Starting at $t = 0$, with probability $q_{ii}$ the node *self generates* a new sequence of words with length $\lambda(t) \sim L$. The probability distribution $L(\cdot)$ can be any length distribution that is representative of natural text lengths. This might vary with $t$, but in this work we exclusively use a Poisson distribution for all time steps $t$. Using this length $\lambda(t)$, the sequence of given length is generated by drawing from any distribution of words or sequences. Each word in the generated sequence shares the same time step index $t$, which will become useful later when we want to draw only from the past of a node. These generated sequences and their associated time indices are concatenated into a single long timestamped sequence for each node.
2. Alternatively, at each time point from $t = 1$ onwards the node can perform a *quote process*. With probability $q_{ij}$, the target can quote from a source, $j$, by selecting a point in the source's sequence uniform randomly. The source's sequence includes only words generated before time $t$. Starting from the selected point in the source's history, a subsequence of length $\lambda(t) \sim L$ is copied from the source's history into the present of the target, and is added to the target's timestamped sequence with time $t$.

Importantly, the quote process draws from the source's past indiscriminately and could copy a sequence of text that the source had previously quoted from elsewhere. This sequence itself could originally have been generated by the target and arrived in the source's history through a series of random quoting rounds. In general, this results in a system of information being passed between the nodes in the network, dependent on the quoting probabilities over the edges.

The model is a dramatic simplification of real social media and news processes but is a useful conceptual model, because we know the underlying network and thus can create accurate measures of the fidelity of an information-flow metric. We can therefore use it to assess whether a metric can realistically estimate the flows in a news ecosystem from the scale of data available, and it can provide relative estimates of which metrics are superior. It is also worth noting that although the model as described above may seem over-simplified, we use real text in the generation component of the model, ensuring that information flows have enough structure to require the more sophisticated Kontoyiannis information rate metrics (as we will show in later comparisons).

## 3. Materials and methods

### 3.1. Flow measures

To better capture the true flow of information in a network a more robust measure is needed. Importantly, information flow measures should have several properties:

1. A metric should be (anti-)symmetric, *i.e.*,

   $flow(A \rightarrow B) = -flow(B \rightarrow A)$,

   in order that net flow between two potential sources is zero. We seek to understand relationships in terms of provider–consumer, and the nature of an asymmetric raw metric like Kontoyiannis' is that it can indicate consumer–consumer relationships that are not helpful in understanding the overall ecosystem.
2. A metric should be real-valued: we seek not just to measure that a relationship exists, but also how strong that relationship is.
3. A metric should not allow a user to enter a network and suddenly appear to contribute significantly to it. If the flow measure was not temporal, it would be possible to create a new user that appeared to contribute a large amount of information purely by copying everything anyone else said; being *first* matters.

Given the above conditions, it is sensible to construct measures of information flow by combining time-synced cross-entropy rates calculated in both directions between nodes.

The first measure we introduce is the simplest, taking the difference between the cross-entropy rates in both directions,

$$\hat{h}(T\|S) - \hat{h}(S\|T). \tag{3}$$

Remembering that the measure needs to identify both the direction and magnitude of the flow across the edge, a positive value for the measure indicates a flow from $S$ to $T$, and the reverse from $T$ to $S$ when the measure is negative.

This measure is un-normalised. If a source was to have complex language (for instance a large vocabulary) leading to high information content, the cross-entropy rate would be naturally inflated regardless of the quoting probability. Normalisation by entropy rate may hence improve detection.

Hence the second and third measures normalise the cross entropy rates by the self-entropy rates of the source,

$$\frac{\hat{h}(T\|S)}{\hat{h}(S)} - \frac{\hat{h}(S\|T)}{\hat{h}(T)}, \tag{4}$$

and the target,

$$\frac{\hat{h}(T\|S)}{\hat{h}(T)} - \frac{\hat{h}(S\|T)}{\hat{h}(S)}. \tag{5}$$

In theory, these self entropy rates may help create a fair comparison between the cross-entropy rates in each direction — as the ability to encode information in a target is affected by the complexity of both the source and the target. We named these the source self normalised information flow (SSNIF) and the self normalised information flow (SNIF).

The final set of measures seek to solve the problem of entropy normalisation, as well as the challenges of increased complexity of quote dynamics (such as the presence of quote cycles and chains of quoting) posed by larger networks. The measures seek to normalise the cross entropy rates using local neighbourhood network information, by dividing by the average cross entropy out of the source,

$$\frac{\hat{h}(T||S)}{\sum_X \hat{h}(X||S)} - \frac{\hat{h}(S||T)}{\sum_X \hat{h}(X||T)}, \quad (6)$$

or into the target,

$$\frac{\hat{h}(T||S)}{\sum_X \hat{h}(T||X)} - \frac{\hat{h}(S||T)}{\sum_X \hat{h}(S||X)}. \quad (7)$$

Similar to above, we name these the source neighbourhood normalised information flow (SNNIF) and the neighbourhood normalised information flow (NNIF). These metrics seek to solve a deeper challenge, namely that in densely connected networks, there can be feedback loops and chains of information flow. Normalising using local network information may provide additional insight into the flow on single edges within the larger network.

In addition we introduce two baseline measures: the first measure uses the time-synced cross entropy measure but takes only the smallest the directed cross-entropy rates,

$$\min(\hat{h}(T||S), \hat{h}(S||T)). \quad (8)$$

One might imagine that this avoids some elements of noise in the less important directional signal.

The second baseline measure uses the Shannon cross-entropy rate between the frequency distribution of the words in each node content,

$$\hat{h}(p, q) - \hat{h}(q, p), \quad (9)$$

where $\hat{h}(p, q) = -\mathbb{E}_p \log q$, and $p$ and $q$ are the probability distributions of the words in the source and target respectively and $\mathbb{E}_p$ is the expected value with respect to $p$. This metric is used primarily as a strawman to show the importance of using the more sophisticated Kontoyiannis-derived information-rate metrics in this (or any other textual) analysis.

### 3.2. News data

This study uses three datasets derived from Twitter for analysis. The first is a large corpus of online new-media tweets. A list of major online news-media organisations was collected from the news-media analysis site AllSides.[1] In this list, each organisation was labelled by political bias as left, lean left, centre, lean right or right using AllSides' rating system [30].

Every tweet over the 2019 calendar year was collected for all organisations from this list with Twitter accounts followed by more than 50,000 people using the Twitter API.[2] Estimation requires a significant volume of data, so we restricted attention to organisations that tweet more than 10 tweets per day on average. This collection includes 2,846,284 tweets over 123 major news-media organisations. Twitter data was chosen for its open availability and representativeness of the stories that news organisations are actively trying to promote. This data mitigates ethical issues with the collection of open textual data from users of social media as it has been produced publicly for the explicit observation by large groups of people. This full dataset and the computed quantities required to reproduce this work are available via Figshare [31] (see Table 1).

---

[1] www.allsides.com
[2] https://developer.Twitter.com/en/docs

**Table 1**
The number of organisations of each bias and the sum of all of their tweets and follower counts.

| Bias | Count | Total tweets | Total followers |
|---|---|---|---|
| Centre | 32 | 938613 | 149186122 |
| Lean left | 34 | 798057 | 165407942 |
| Lean right | 12 | 228644 | 4958618 |
| Left | 29 | 571974 | 97645740 |
| Right | 16 | 308996 | 8484882 |

To supplement the Twitter data, we also collect the news story text from the 20 most followed news organisations from the dataset. Three thematically-varying major news events throughout 2019 were selected: the New Zealand Christchurch massacre, the college admissions bribery scandal, and Greta Thunberg's selection as TIME's Person of the Year. These were chosen for the popularity of the tweets linking to news stories and the time-evolving nature of these stories which resulted in more textual data (needed for estimation to converge) and more complex information dynamics (beyond a single organisation sharing a single fact).

For each news event, we manually select a broad set of keywords which we use to filter tweets from the most followed news organisations. The article links from these tweets are followed and the news article text's are scraped. The limited scope of news events and accounts allow us to overcome the slow and unreliable nature of web scraping as a collection tool. For each story, a more constrained set of related keywords is used to check the relevance of a story, and a final manual inspection is performed to verify that collected articles from these major outlets correspond to the correct news event. Since publishing dates are often more difficult to scrape, we associate the earliest tweet of an article as its publishing date. This collected text corpora is available on Figshare [32].

Finally, our third dataset uses the roughly 3 million Russian troll tweets released by FiveThirtyEight [33] collected during the 2016 election. This dataset, although not news itself, involved bot sharing informational (or moreover disinformational) content, and is set in a different time period to help contrast with the above data and demonstrate the broader applicability of these information flow methods. We use the FiveThirtyEight over the original Twitter data for the convenience of the labels defining each account into five different troll categories.

#### 3.2.1. Text generation

In the quoter model nodes in the network can either quote from adjacent nodes or *self-generate* text. This self-generation process can use any arbitrary method for generating text at each time point. In the case of the original quoter model, two methods are used; drawing uniformly from a fixed vocabulary and drawing from a rank-ordered vocabulary according to a Zipf distribution. Here we build our text generation using the real data in two ways. Our first approach draws real tweets uniformly and without replacement from the pool of tweets by all news-media organisations. This provides real natural language text from which to induce quoting and draw uniformly from all sources helps average out any possible underlying information flows between organisations.

The second approach uses a Zipf law fitted to the data. As has been seen in other corpora [34,35], the rank-frequency distribution of the text in the news tweet corpus exhibits two different scaling regimes; common words scale with $\alpha \approx 0.8$ and uncommon words scale with a much higher $\alpha \approx 2$, resulting in an overall scaling parameter of $\alpha = 1.2$ which we use here to generate synthetic text.

*3.3. Network experiments*

We use quoter model simulations on larger networks to examine how the various measures perform under various network conditions. Two such experiments are performed: a measure performance evaluation on a large sample of networks with varied parameters and a sensitivity analysis on networks with only one parameter changing.

To compare performance across conditions we generate ER(*n*, *p*) quoter networks over a grid of parameters, $n \in [10, 40]$, $p \in [0.08, 1]$. Two such experiments are performed, one in which nodes self-generate using real text data as above and one in which nodes self-generate using synthetic text data.

In both cases the quoting process based on randomly selected edge weights is identical. Self-generation and quoting occurs for 7500 time steps to ensure that sufficient textual data is present to ensure the cross-entropy rate estimator converges fully. The measures are evaluated on the Pearson correlation coefficient $r$ and Spearman correlation coefficient $\rho$ between the induced quote probabilities and the information flow measure. An accuracy score is determined by the number of edges in which the information flow was determined to be in the correct direction of the induced quoting.

To test the sensitivity, four experiments are run. The first simulates the quoter model as above over a clique with each edge randomly assigned a direction and U(0,1) weight with clique sizes ranging from 2 to 50. The self-generation probability is held constant at 0.5 and the quote probabilities are normalised with this to sum to one. Networks are repeatedly generated for each network size with 500 networks for size 2 and 4 networks for size 50 to ensure that sufficient edges and network structures are generated to provide confidence in the Pearson correlation coefficients calculated for each size.

The second experiment is similar to the first but fixes the clique size at 20 nodes and varies the self-generation probability. The main effect of this is reducing the average quote probabilities since quote probabilities are normalised with the self-generation probability to sum to 1. This allows for the effect of network size to be disambiguated from the effect of quote probabilities.

Following from this, the third experiment uses a Watts–Strogatz [36] network with 20 nodes and starting degree 4. We vary the rewiring parameter $\beta$ to examine the effect of network density on the information flow measures.

Finally, an ER(20, *p*) quoter model network is simulated for varying *p* and a fixed self-generation probability of 0.5 to examine the effect of changing quote probabilities in networks of different densities. Across these experiments networks are repeatedly generated such that *at least* 7500 edges are estimated over.

*3.4. Applications to real data*

Using the results from the simulated network experiments, we identify the best information flow measure and apply it to the three different datasets.

We first apply the flow estimation to all pairs of news-media organisation using their Twitter timelines. In doing so we seek to answer what organisations, on average, are contributing the most information into the ecosystem and to assess if there are structural trends such as the dominance of major national news organisations.

Secondly, we take the linked full-text news article data and run a similar information flow analysis between each pair of news organisations. The constrained scope of these stories allows for a qualitative examination of the estimated flows to identify what information trends are observed and how these effect the estimated flow quantities. To contrast outlying information flows in this application, we will use all pairwise information flows for the news event as the distribution from which to compare large flows and compute z-scores.

Finally, we repeat this analysis on a third dataset which contains 3 million tweets from accounts associated with the Internet Research Agency, a known Russian "troll factory" [33]. Each account in this dataset is labelled, and we assess whether the median net information flow measure is significantly different between each labelled account type. This is achieved by calculating an empirical *p*-value by permuting the labels on the dataset, a method which can establish statistical significance without assumptions of normality [37].

**Table 2**
The performance scores for each measure of information flow in ER(*n*, *p*) quoter model networks with a grid a parameters $n \in [10, 40]$, $p \in [0.08, 1]$ using both real and synthetic text data.

| Measure | Synthetic | | | Real | | |
|---|---|---|---|---|---|---|
| | *r* | % acc | $\rho$ | *r* | % acc | $\rho$ |
| $\hat{h}(T\|\|S) - \hat{h}(S\|\|T)$ | 0.859 | 0.598 | 0.762 | 0.864 | 0.584 | 0.764 |
| $\frac{\hat{h}(T\|\|S)}{\hat{h}(S)} - \frac{\hat{h}(S\|\|T)}{\hat{h}(T)}$ | 0.636 | 0.512 | 0.522 | 0.642 | 0.498 | 0.520 |
| $\frac{\hat{h}(T\|\|S)}{\hat{h}(T)} - \frac{\hat{h}(S\|\|T)}{\hat{h}(S)}$ | 0.940 | 0.655 | 0.887 | 0.938 | 0.638 | 0.878 |
| $\frac{\hat{h}(T\|\|S)}{\sum_X \hat{h}(X\|\|S)} - \frac{\hat{h}(S\|\|T)}{\sum_X \hat{h}(X\|\|T)}$ | 0.839 | 0.587 | 0.737 | 0.840 | 0.577 | 0.741 |
| $\frac{\hat{h}(T\|\|S)}{\sum_X \hat{h}(T\|\|X)} - \frac{\hat{h}(S\|\|T)}{\sum_X \hat{h}(S\|\|X)}$ | **0.969** | **0.668** | **0.925** | **0.968** | **0.654** | **0.918** |
| $\min(\hat{h}(T\|\|S), \hat{h}(S\|\|T))$ | 0.519 | 0.598 | 0.309 | 0.510 | 0.584 | 0.302 |
| $\hat{h}(p, q) - \hat{h}(q, p)$ | 0.048 | 0.358 | 0.003 | 0.057 | 0.345 | 0.005 |

**4. Results**

The performance comparison between measures shows a clear separation of measures between normalisation methods. Across all three performance measures in both the real and synthetic text networks in Fig. 2 the relative performance rankings are the same between measures. The highest performing measure is the neighbourhood normalised information flow (NNIF), $\frac{\hat{h}(T\|\|S)}{\sum_X \hat{h}(T\|\|X)} - \frac{\hat{h}(S\|\|T)}{\sum_X \hat{h}(S\|\|X)}$, while the second-highest performing measure is normalised only by the entropy rate calculated on the target without comparison to other nodes, the self normalised information flow (SNIF), $\frac{\hat{h}(T\|\|S)}{\hat{h}(T)} - \frac{\hat{h}(S\|\|T)}{\hat{h}(S)}$.

These measures perform better than the measure without any normalisation, which itself surprisingly performs better than the measure which normalised by the entropy rate or neighbourhood cross-entropy rate of the *source*.

The baseline measures underperform, with the minimum of the cross-entropy rates measure demonstrating that there is a notable increase in performance by taking entropy rate difference rather than using a single cross-entropy calculation. The baseline measure using Shannon entropy can accurately identify the direction of information flow 40% of the time is largely unable to estimate the magnitude of that flow. The ability to identify direction can largely be attributed to the fact that increasing quoting rates will decrease the vocabulary size of the target compared to the source as repeated random sampling from a finite source will result in smaller state spaces.

Of additional note are the similarity in both the synthetic text generation and real text data performance. The coherence of the results help verify that the possible flaws in text generation – existing bias from real text or a lack of realism in synthetic text generation – are not effecting the analysis, as demonstrated by complementary trade-offs in the approaches (see Table 2).

*4.1. Sensitivity analysis*

These findings hold in the sensitivity analysis which compares the performance across different network conditions. Fig. 3(a) demonstrates that increasing size on networks makes information flow estimation more difficult. There are slight declines in performance for the NNIF and SNIF while other measures dramatically reduce in performance for large networks. The NNIF measure scales best, but for networks with less than 4 nodes, SNIF performs better as the small neighbourhood of nodes is not large enough to normalise. While ever larger networks may reduce performance further, computational time on the simulations to achieve significance limits this analysis and is left to future work.

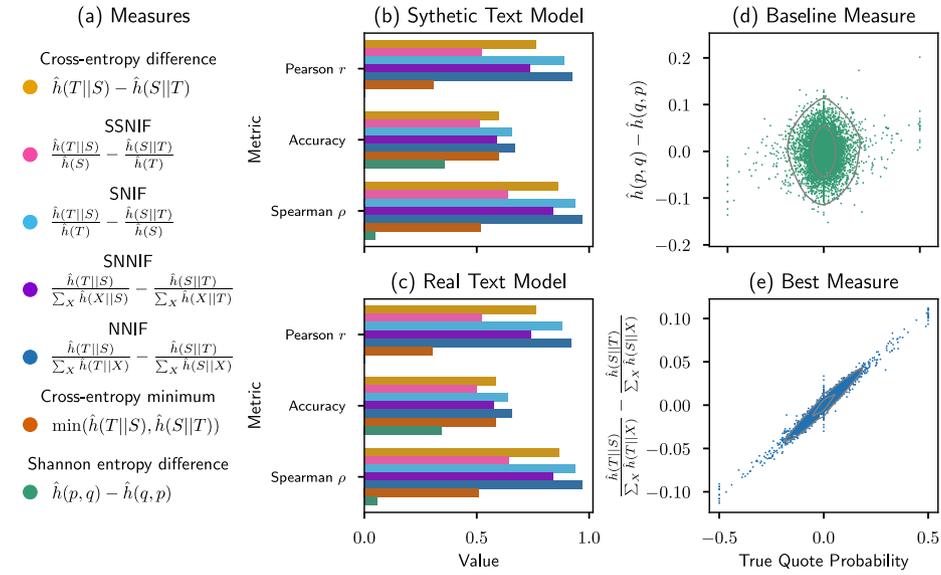

**Fig. 2.** Performance comparison of information flow measures. Networks are reconstructed from these quoter models and the performance of the measures are compared for networks of synthetic text data (b) and real text data taken from news organisations (c). Measures that normalise by the entropy rate of a target (e) perform significantly better than baseline measures (d) and those that do not. ER($n$, $p$) quoter networks are repeatedly generated with a grid a parameters $n \in [10, 40]$, $p \in [0.08, 1]$ to provide confidence in estimates.

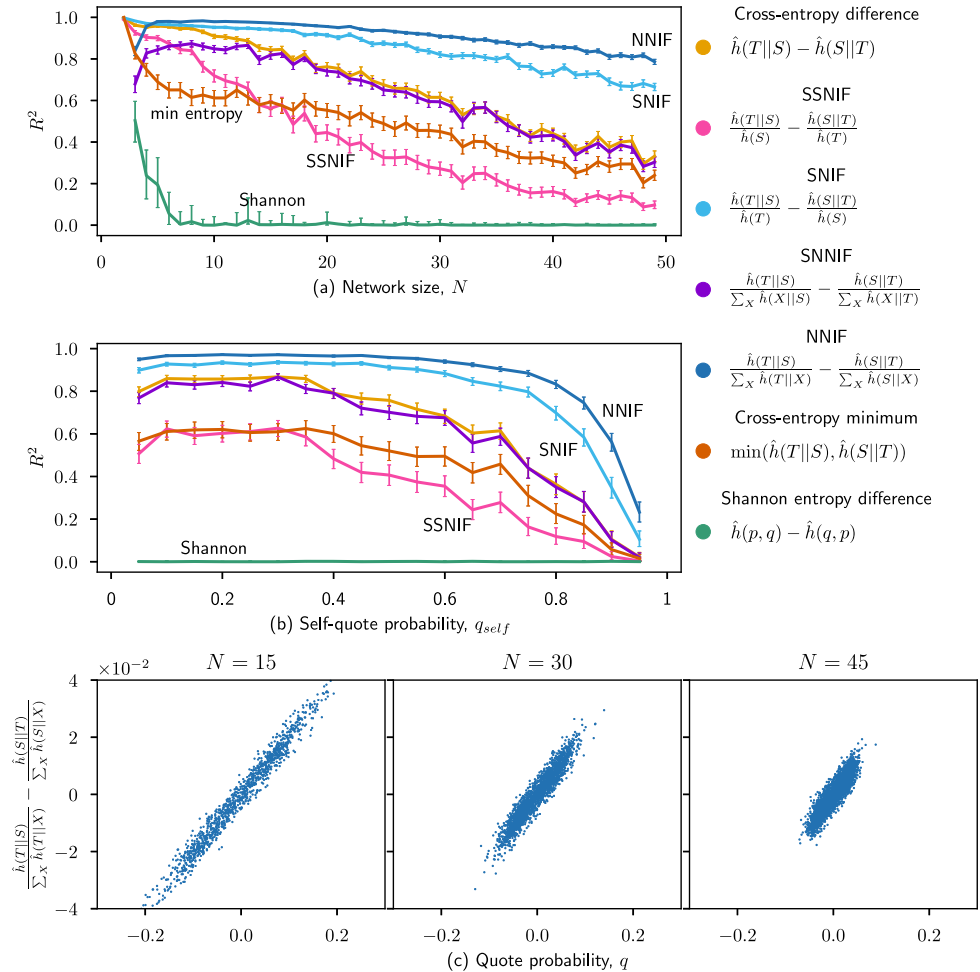

**Fig. 3.** Sensitivity analysis of the information flow measures. Quoter model networks are simulated in (a) on cliques of increasing size using synthetic data and all pairwise edges connected with a direction and weight chosen randomly before normalisation. Increasing networks size decreases measure performance only slightly for the neighbourhood normalised information flow (NNIF) and self normalised information flow (SNIF) measures. While rankings are consistent, (b) shows that in a size 20 clique increasing self-generation probabilities, and hence decreasing normalised edge quote probabilities, reduces the performance significantly as the ratio of the signal from quoting to the natural noise of textual information decreases. A similar effect can be seen to justify the reduction in network size performance in (c).

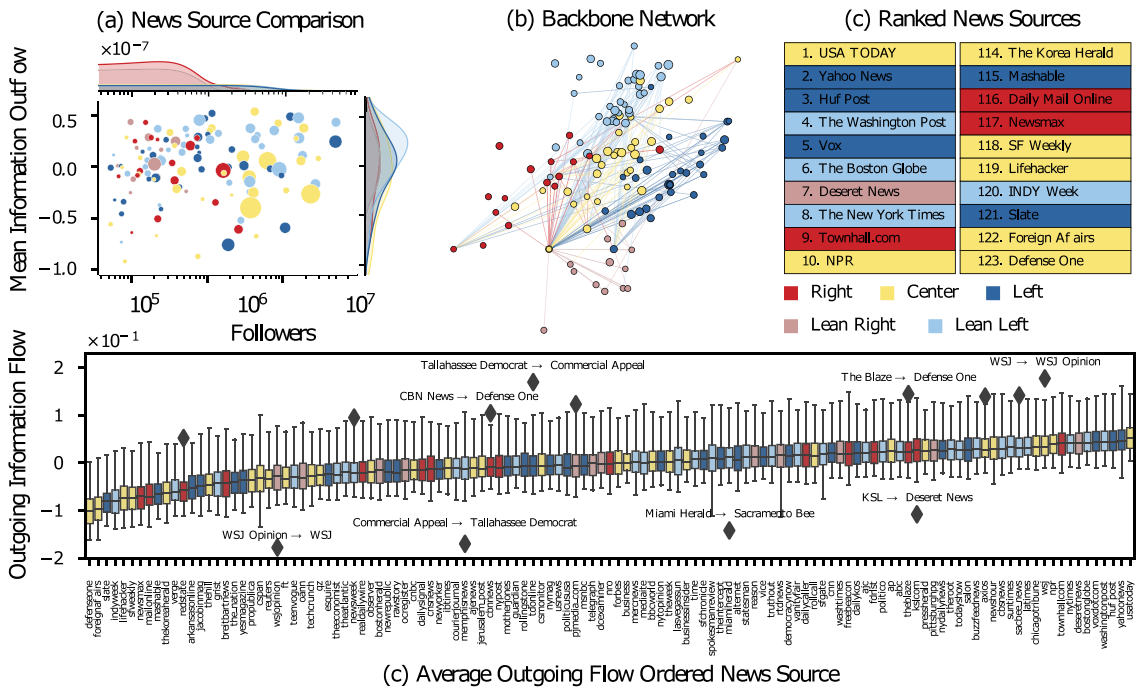

**Fig. 4.** NNIF is applied to each pair of news media organisations. The mean outgoing information flow is compared to the number of followers and the total number of tweets (shown as size) in (a), no significant correlation is found between follower count or number of tweets and the average outgoing flow. (c) shows the distribution of outgoing information flow for each news organisation and outliers are identified as news organisations owned by the same network. These average information outflows are used to rank the news sources by information contribution in (c) and a backbone network where edges are weighted directed flows is shown in (b).

As network size increases for these cliques, quote probabilities which are normalised to 1 in each neighbourhood get smaller and hence the signal to noise ratio between the quoting and underlying textual information noise decreases as in Fig. 3(c). To distinguish between the effect of the increasing network complexity and decreasing quote probabilities, we examine the effect of changing the self-generation probability in the 20 node clique of Fig. 3(b). At self-generation probabilities, $q_{self}$ reach above 0.8, the performance of the measures drops significantly. Edge quote probabilities, are on average, $(1 - q_{self})/20$ which gives an average quote probability of 0.01 above $q_{self} = 0.8$. As textual information has an underlying noise by the randomness with which we quote similar expressions, measures face difficulty identifying such small information flows.

Supporting this result, the Erdős–Rényi experiment showed that increasing the number of edges in the fixed-size graph resulted in a small decline in performance while the Watts–Strogatz experiments shown no effect of rewiring on measure performance. In both additional experiments, the relative rankings of the measures remained unchanged.

### 4.2. Application to tweets from news organisations

The application of these measures to the real data yields interesting results. Firstly, by estimating the information flow across Twitter feeds for all pairs of news organisations we can find which pairs of organisations have the greatest information flow. The largest flow is from the Wall Street Journal to its opinion column WSJ Opinion. Similarly, the second largest information flow is from the Tallahassee Democrat to Commercial Appeal (@memphisnews) which are sister channels in the USA Today Network. In both of these cases the large net flow indicates not only the volume of duplicate posted but also which account generally posts first. Both of these flows appear as outliers in the source outflow boxplots in Fig. 4(c). In contrast, the third-largest flow is not an outlier, with a large flow from Yahoo News to Defence One. YahooNews has a high average outflow as it posts broadly relevant information early and often. Defence One is a significant net information absorber, due to at least two effects: it is less active than many organisations and has a niche subject focus. The latter matters as very few other organisations are as active on the subject of defence, and hence less information flows out of Defence One.

Beyond individual flows, we can rank the news organisations net influence on the network by their average outgoing flow. In Fig. 4(c) we see such a ranking where, somewhat unsurprisingly, we see several of the largest U.S. news organisations (USA Today, Washington Post, New York Times). However among the top ten are several smaller organisations, most notably Deseret News, which although a small organisation, was a strong net contributor to the information ecosystem by producing information that was timely and somewhat relevant to the broader network while absorbing very little information from larger sources. Only KSL News – the other Utah based news organisation – was a significant information contributor to Deseret.

More broadly, Fig. 4(a) shows the relationships between the numbers of followers, the total number of tweets and average information outflow for each news Twitter account. We find no significant correlation between the number of followers or the number of tweets and the average outflow. The right and right-leaning accounts have a much smaller number of Twitter followers on average, and yet still contribute a large amount of information into the news ecosystems.

### 4.3. Application to full text articles

In addition to analysing the tweets from the major news sources, we analyse the news article text itself for a small set of stories present in top news sources.

In the case of the Greta Thunberg Time Person of the Year story, Time Magazine had an above-average – although not significant ($p =$

(a) Net information flow out of accounts grouped by source type.      (b) Net information flow out of accounts by source type.

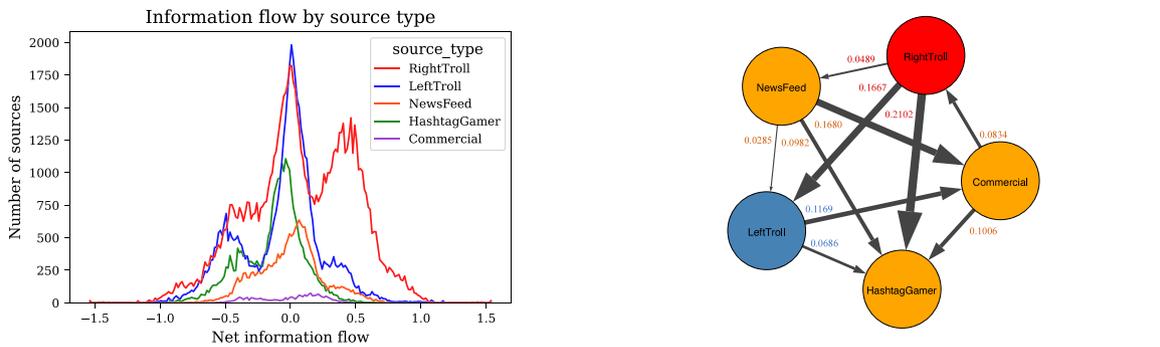

(c) Empirical Null Distribution of median net flow between groups.

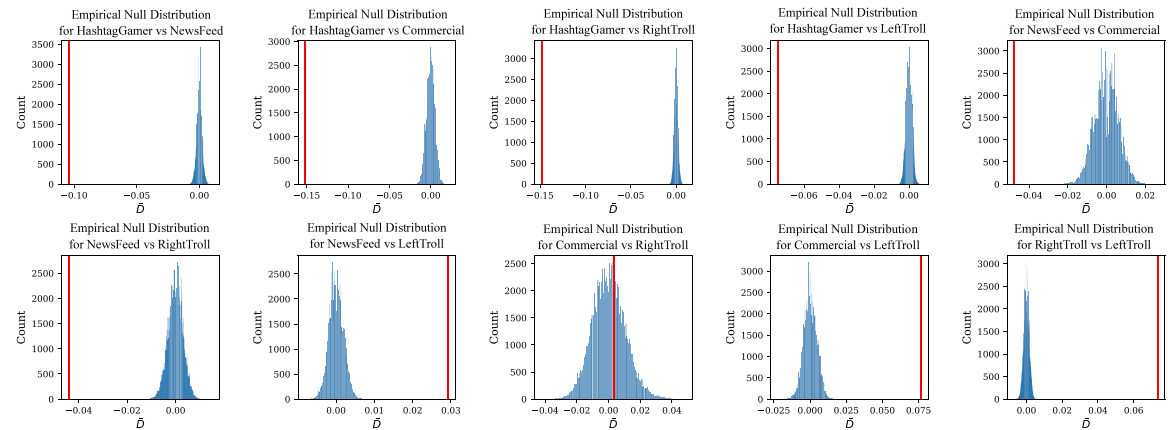

**Fig. 5.** Analysis of net information flow measure by account group as a metric to distinguish and characterise activity of malicious Russian trolls on Twitter. The distribution of outward net information flow by group type (a) indicates this measure captures distinct behaviour types between groups. Most strikingly, the net information flow from the accounts labelled `RightTroll` seems to dominate the outward information flow. This is apparent in the network created by considering the mean information flow by account type (b) which shows there is a strong flow of information from accounts labelled `RightTroll` into accounts labelled `LeftTroll` and `HashtagGamer`. We also see a large flow from `NewsFeed` labelled accounts to accounts labelled `Commercial`. (c) Empirical null distribution of the difference in medians of the net information flow between groups calculated using 100,000 samples. Let $\hat{D}$ be the difference in median net information flow between the two groups. Shown in red are the observed difference between the medians of the two groups. An empirical *p*-value was created for the difference of median information flow between each labelled group, showing a significant difference for all pairs except the `Commercial` and `RightTroll` groups. (For interpretation of the references to colour in this figure legend, the reader is referred to the web version of this article.)

.269) – outgoing information flow to all sources. However, the flow network reveals a more interesting progression of information flow. Vanity Fair had an average larger outgoing information flow to most sources while having a large and significant flow from Time to Vanity Fair ($p = .0135$). This network reveals that while Time produced the original story, Vanity Fair propagated the "Greta Thunberg says she 'Wouldn't have wasted my time' talking climate change with Trump" framing, bringing discussions of Trump into the story information ecosystem. What this progression highlights is that the flow estimation to the whole network from original sources can be dampened as other sources riff on the story, adding new framing that becomes more popular, and hence induces more substring matching and lower entropy calculations.

In contrast, the Christchurch mosque shootings story had no single source that had a significantly higher than average of outgoing information flow. Individual organisations would produce articles before one another based on resource constraints and publication times, but as the global discourse went on, the balance of who contributed what information and who else copied said information shifted between different outlets. This is strongly driven by the nature of the story, where a large amount of novel information originated from the New Zealand government (who's press releases are not included in the analysis), which all the other news organisations drew from.

A similar conclusion can be seen in the Operation Varsity Blues college admission scandal, where the no clear originator of most information exists, since public FBI press releases and other sources are available to all the news organisations. While some flow edges are significantly larger than rest of the edge weight distribution – such as HuffPost to USA Today (z-score of 2.73) – these can be explained by the order of the articles being published and strong overlap in their language use.

### 4.4. Application to inauthentic Russian troll activity

On the third dataset of 3 million tweets from a known Russian "troll factory" [33] that was active during the 2016 US presidential election, the net information flow reveals differences in the behaviour of different account types. Each account in this dataset is classified into an account type using account and content level features, including account name and behaviour. The naming of the account classification labels aims to represent the authentic user groups the accounts are attempting to mimic.

These accounts work together to form a network, building credibility for their own agenda around topics ranging from the 2016 US political election to social movements such as the Black Lives Matter movement [38–40]. Applying the net information flow between every pair of troll accounts and then grouping by types reveals patterns in the flow of information amongst this network, with large volumes of information flowing out of account types with the label `RightTroll` and `NewsFeed`, shown in Fig. 5. From this we can observe that these account types are driving the discussions and ideas amongst this network, implying that while these accounts exhibit a range of different behaviours, there is an overall structure in the process these malicious campaigns followed. That the overall net information flow is greatest from the `RightTroll` group is interesting, suggesting a deliberate strategy was being deployed by the troll factory to imitate left-leaning accounts responding to new content being broadcast by right-leaning accounts. Whether this behaviour is similarly observed in authentic political discussion is a question for future research; as is a deeper understanding of the underlying social influence strategy being deployed that these empirical results are suggestive of.

To support this, we consider the median net information flow between groups, using an empirical *p*-value to establish significant differences between these values in all accounts pairings except with the `Commercial` and `RightTroll` group accounts. This result shows that the net information flow measure is revealing significant and useful insight into the mechanisms which are driving information flow across this network, and supports account labelling performed using content and account level metrics independent of information flow.

## 5. Discussion

Our results show that even in interconnected systems of natural language text, estimates of temporal net information flow can be reliably produced given enough textual data. For large interconnected systems, the local neighbourhood measure, NNIF, should be used as its performance scales well with size and small quoting probabilities. Where there are less than 4 nodes to compare, normalising by self entropy rates, SNIF, is preferable.

Normalising with self entropy rates instead of local neighbour information does scale similarly to the NNIF measure but requires the calculation of self entropy rates for all nodes in the network. This is an additional computational expense on top of what is already a time-consuming process. High-efficiency code has been developed and is available in the ProcessEntropy package on the Python Package Index.[3] This package utilises compiled machine code to speed up NumPy functions in Python, although the algorithm is limited by the time complexity of the pairwise cross entropy estimation (which all measures require) of

$$O\left(N^2 T \log^2 T + N^2(T+L)\log T\right), \qquad (10)$$

with $N$ nodes containing an average of $T$ tokens and with $L$ as the average sum of longest common substring lengths. This results in compute times that take from a few hours to a few days depending on compute resources using data of roughly the size here.

It is worth noting that both NNIF and SNIF normalise by entropy rates of a target (the destination of information flow). At its core, the cross-entropy calculation is designed to describe the complexity of the target given the social data obtained from the source — it represents the number bits per word required to describe the target using information from the source.

While these results suggest that information flow can be accurately extracted, this approach has several limitations.

- Twitter is an inherently challenging data source. Its use by news organisations varies and tweets often show only a 'click-bait' title or key finding of an article. Moreover, timing is important in ascribing the origin of a piece of information, and the time at which news organisations post on Twitter is only a proxy for the time that an organisation releases news.
- The complexity of our cross-entropy rate estimation is polynomial, but large with respect to the length of text estimated. It scales poorly to extremely large text corpora. Indeed, the analysis here took a large amount of compute to produce and can be difficult to reproduce on larger networks with more content. Future and currently ongoing work aims to reduce the computational complexity through use of prefix-tree string matching algorithms and more resources dedicated to refining the code base.

---

[3] https://pypi.org/project/ProcessEntropy/

## 6. Conclusion

This work shows that a useful measure of net information flow can be defined in systems of natural language text by calculating the cross-entropy rate using a time-synced nonparametric estimator. This work demonstrates that these cross-entropy rate estimates are insufficient to identify information flow unless normalisation is performed. This flow can either be estimated by the network normalised information flow (NNIF) measure or the self normalised information flow (SNIF) measure. Through simulation experiments we find that NNIF is preferred for systems with larger than four nodes in the network. These measures perform well under a variety of network conditions, and lose performance when information flow is small enough to be indistinguishable from the "noise" of natural language. The methods presented here apply to other systems in social media, emails or the web and require only a system of text producers sharing information and creating timestamped text, subject to the scalability of the computational tools used here.

The NNIF is applied to a large corpus of news organisations on Twitter and rankings are estimated based on the average net outgoing information flow from each news source. We find several of the largest U.S. news organisations top the influence rankings but that no significant correlation is found between the organisations number of followers or number of tweets and its ranking. The right and right-leaning organisations have fewer followers on average and yet still contribute a large amount of information into the news ecosystems. Similarly, the methods are applied to smaller datasets including the full-text of news articles on a range of topics from 2019 and on a large dataset of 3 million Russian troll accounts. In both cases the information flow measure allow for the drivers of the information narratives to be identified and quantified, an important task in ascribing journalistic credit and combating disinformation.

## CRediT authorship contribution statement

**Tobin South:** Conceptualization, Methodology, Software, Data curation, Investigation, Formal analysis, Writing – original draft. **Bridget Smart:** Visualization, Investigation, Validation, Writing – review & editing. **Matthew Roughan:** Writing – review & editing, Supervision, Resources. **Lewis Mitchell:** Conceptualization, Writing – review & editing, Supervision, Resources.

## Declaration of competing interest

The authors declare that they have no known competing financial interests or personal relationships that could have appeared to influence the work reported in this paper.

## Acknowledgements

TS acknowledges a Fulbright Future Scholarship, and BS acknowledges a Westpac Future Leaders Scholarship. LM and MR are supported by the Australian Government through the Australian Research Council's Discovery Projects funding scheme (project DP210103700).